\documentclass[letterpaper, 10 pt, conference]{ieeeconf}

\usepackage{amsfonts,amsmath,amssymb} 

\def\rbb{\mathbb{R}}
\def\trp{^T}

\def\half{\frac{1}{2}}
\usepackage{psfrag,color}
\usepackage{enumerate,cite,latexsym,graphicx}
\newtheorem{theorem}{Theorem}

\newtheorem{definition}{Definition}

\newtheorem{remark}{Remark}

\title{\LARGE \bf A Direct Coupling Coherent Quantum Observer for a Single Qubit Finite Level Quantum System}

\author{Ian R.~Petersen %
\thanks{This work was supported by the
Australian Research Council (ARC) and the Air Force Office of Scientific
Research (AFOSR). This material is based on research sponsored by the
Air Force Research Laboratory, under agreement number FA2386-12-1-4075.  The U.S. Government is authorized to reproduce and
distribute reprints for Governmental purposes notwithstanding any
copyright notation thereon.
The views and conclusions contained herein are those of the authors
and should not be interpreted as necessarily representing the official
policies or endorsements, either expressed or implied, of the Air
Force Research Laboratory or the U.S. Government. }%
\thanks{Ian R. Petersen is with the School of  Engineering and Information Technology, 
        University of New South Wales at the Australian Defence Force Academy, Canberra ACT 2600, Australia.
         {\tt\small i.r.petersen@gmail.com} } 
}%

\def\begce{\begin{center}}
\def\endce{\end{center}}
\def\begar{\begin{array}}
\def\endar{\end{array}}
\def\begeq{\begin{equation}}
\def\endeq{\end{equation}}
\def\begdi{\begin{displaymath}}
\def\enddi{\end{displaymath}}
\def\begdis{\begin{eqnarray*}}
\def\enddis{\end{eqnarray*}}
\def\begeqa{\begin{eqnarray}}
\def\endeqa{\end{eqnarray}} 

\def\re{{\mathbb R}}
\def\C{{\mathbb C}} 
        
\begin{document}

\maketitle
\thispagestyle{empty}
\pagestyle{empty}

\begin{abstract}
This paper considers the problem of constructing a direct coupling quantum observer for a single qubit finite level quantum system plant. The proposed observer is a single mode linear quantum system which is shown to be able to estimate one of the plant variables in a time averaged sense. A numerical example and simulations are included to illustrate the properties of the observer.
\end{abstract}

\section{Introduction} \label{sec:intro}
In order to better understand fully quantum
estimation and control, a number recent papers have introduced a class of coherent quantum observers  for linear
quantum stochastic systems; see \cite{MJ12a,VP9a}.  Also, the paper \cite{EMPUJ6a} considers a
finite level quantum system as the quantum plant, which is described in the form of bilinear quantum stochastic
differential equations (QSDEs); see \cite{EMPUJ1a,EMPUJ2a,EMPUJ3a,EMPUJ4a}. This means that the combined plant observer system is a hybrid of a finite level quantum system and a linear quantum system; see \cite{EMPUJ5a}. 

The coherent observers discussed in \cite{MJ12a,VP9a,EMPUJ6a} track the plant variables asymptotically in the sense of mean
values. Also, entanglement can be generated in the joint plant-observer
quantum systems \cite{MJ12a}. 

In the papers  \cite{MJ12a,VP9a}, the quantum plant under consideration is a linear quantum system. In recent years, there has been considerable interest in the modeling and feedback control of linear quantum systems; e.g., see \cite{JNP1,NJP1,ShP5}.
Such linear quantum systems commonly arise in the area of quantum optics; e.g., see
\cite{GZ00,BR04}. For such linear quantum system models an important class of quantum control problems are referred to as coherent
quantum feedback control problems; e.g., see \cite{JNP1,NJP1,MaP3,MAB08,ZJ11,VP4,VP5a,HM12}. In these coherent quantum feedback control problems, both the plant and the controller are quantum systems. The coherent quantum observer problem can be regarded as a special case of the coherent
quantum feedback control problem in which the objective of the observer is track the system variables of the quantum plant. 

\begin{figure}[htbp]
\begin{center}
\includegraphics[width=8cm]{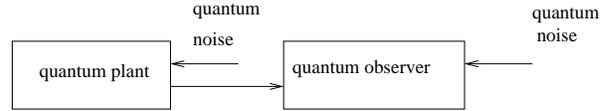}
\end{center}
\caption{Coherent Observer Structure with Field Coupling.}
\label{F1}
\end{figure}

In the previous papers on quantum observers such as  \cite{MJ12a,VP9a,EMPUJ6a}, the coupling between the plant and the observer is via a field coupling. This leads to an observer structure of the form shown in Figure \ref{F1}. This enables a one way connection between the quantum plant and the quantum observer. Also, since both the quantum plant and the quantum observer are  open quantum systems, they are both subject to quantum noise.  

\begin{figure}[htbp]
\begin{center}
\includegraphics[width=8cm]{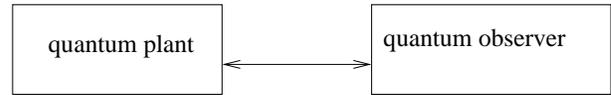}
\end{center}
\caption{Coherent Observer Structure with Direct Coupling.}
\label{F2}
\end{figure}

In the paper \cite{ZJ11}, a coherent quantum control problem is considered in which both field coupling and direct coupling is considered between the quantum plant and the quantum controller. Also, the paper \cite{PET14Aa} considered a direct coupling quantum observer in which there is only direct coupling between the quantum plant and the quantum observer and for which both the plant and the observer are linear quantum systems corresponding to quantum harmonic oscillators. In this paper, we consider the construction of a coherent quantum observer in which there is only direct coupling between quantum plant and the quantum observer. Also, the plant is assumed to be a finite level quantum system corresponding to a single qubit and the observer is assumed to be a linear quantum system corresponding to a single quantum harmonic oscillator. Furthermore, both the quantum plant and the quantum observer are assumed to be closed quantum systems which means that they are not subject to quantum noise and are purely deterministic systems. This leads to an observer structure of the form shown in Figure \ref{F2}. It is shown that for the case being considered, a quantum observer can be constructed to estimate one of the system variables of the quantum plant. In particular, an observer variable converges to the plant variable being estimated in a time averaged sense.

\section{Quantum  Systems}
We first consider  the dynamics of a single qubit spin system which will correspond to the quantum plant; see also \cite{EMPUJ1a,EMPUJ2a}.  
 The quantum mechanical behavior of the system is described in terms of the system \emph{observables} which are self-adjoint operators on the complex Hilbert space $\mathfrak{H}_p = \C^2$.   The commutator of two scalar operators $x$ and $y$ in ${\mathfrak{H}_p}$ is  defined as $[x, y] = xy - yx$.~Also, for a  vector of operators $x$ in ${\mathfrak H}_p$, the commutator of ${x}$ and a scalar operator $y$ in ${\mathfrak{H}_p}$ is the  vector of operators $[{x},y] = {x} y - y {x}$, and the commutator of ${x}$ and its adjoint ${x}^\dagger$ is the  matrix of operators 
\begdi [{x},{x}^\dagger] \triangleq {x} {x}^\dagger - ({x}^\# {x}^T)^T, \enddi 
where ${x}^\# \triangleq (x_1^\ast\; x_2^\ast \;\cdots\; x_n^\ast )^T$ and $^\ast$ denotes the operator adjoint. In the case of complex vectors (matrices) $^\ast$ denotes the complex conjugate while $^\dagger$ denotes the conjugate transpose. 

The vector of system variables for the single qubit spin system under consideration is 
\begdi x_p=(x_1,x_2,x_3)^T\triangleq (\sigma_1,\sigma_2,\sigma_3),\enddi 
where $\sigma_1$, $\sigma_2$ and $\sigma_3$ are spin operators. Here, $x_p$ a self-adjoint vector of operators, i.e., $x_p=x_p^\#$.~In particular $x_p(0)$ is represented by the Pauli matrices; i.e.,
\begin{eqnarray*}
\sigma_1(0)&=&\left(\begin{array}{cc}
         0 & 1 \\ 1 & 0
        \end{array} \right),\;\; 
\sigma_2(0)=\left(\begin{array}{cc}
         0 & -{\pmb i} \\ {\pmb i} & 0
        \end{array} \right),\\
\sigma_3(0)&=&\left(\begin{array}{cc}
         1 & 0 \\ 0 & -1
        \end{array} \right).
\end{eqnarray*}
 Products of the spin operators satisfy \begdi \sigma_i\sigma_j = \delta_{ij}+ {\pmb i} \sum_{k}\epsilon_{ijk}\sigma_k.\enddi It is then clear that the commutation relations for the spin operators are
\begeq \label{eq:Pauli_CCR}
[\sigma_i,\sigma_j] = 2{\pmb i} \sum_{k}\epsilon_{ijk}\sigma_k,
\endeq
where $\delta_{ij}$ is the Kronecker delta and $\epsilon_{ijk}$ denotes the Levi-Civita tensor. The dynamics of the system variables $x$ are determined by the system Hamiltonian which is a self-adjoint operator on $\mathfrak{H}_p$. The {Hamiltonian} is chosen to be linear in $x_p$; i.e., 
\begdi {\mathcal{H}}_p=r_p^T x_p(0) \;\;\enddi
where $r_p\in \re^3$. 
The plant model is then given by the differential equation
\begin{eqnarray}
\dot x_p(t) &=& -{\pmb i}[x_p(t),\mathcal{H}_p]; \nonumber \\
&=& A_px_p(t); \quad x_p(0)=x_{0p}; \nonumber \\
z_p(t) &=& C_px_p(t)
 \label{plant}
\end{eqnarray}
where $z_p$ denotes the vector of system variables to be estimated by the observer and  $C_p\in \rbb^{1\times 3}$; e.g., see \cite{EMPUJ1a}. Also, $A_p \in \re^{3\times 3}$. In order to obtain an expression for the matrix $A_p$ in terms of $r_p$, we  define the linear mapping
$\Theta: \C^3 \rightarrow \C^{3\times 3}$ as
\begeq \label{eq:Theta_definition}
\Theta(\beta)= \left(\begin{array}{ccc}
         0 & \beta_3 & -\beta_2 \\ -\beta_3 & 0 & \beta_1 \\ \beta_2  & -\beta_1  & 0
        \end{array} \right).
\endeq  
Then, it is shown in \cite{EMPUJ1a} that
\begdi x_p(t)x_p(t)^T=I + {\pmb i} \Theta(x_p(t)). \enddi 
Similarly, the commutation relations for the spin operators are written as 
\begin{equation}
\label{comm_1}
 [x_p(t),x_p(t)^T]=2 {\pmb i} \Theta(x_p(t)). 
\end{equation}
Also, it was shown in \cite{EMPUJ1a} that 
\begin{equation}
-{\pmb i}[x_p(t),r_p^T x_p(t)] = - 2  \Theta(r_p) x_p(t)
\label{Ap}
\end{equation}
and hence $A_p = - 2  \Theta(r_p)$. 

In addition, it is shown in \cite{EMPUJ1a} that the mapping $\Theta(\cdot)$ has the following properties:
\begin{eqnarray}
\label{eq:Theta_1} \Theta(\beta)\gamma &=& - \Theta(\gamma) \beta,\\
 \label{eq:Theta_beta_beta} \Theta(\beta)\beta &=& 0,\\
 \label{eq:Theta_multiplication} \Theta(\beta)\Theta(\gamma) &=& \gamma \beta^T -\beta^T \gamma I,\\
 \label{eq:Theta_composition} \Theta\left(\Theta(\beta)\gamma\right)&=&\Theta(\beta)\Theta(\gamma) - \Theta(\gamma)\Theta(\beta).
\end{eqnarray}

Note that a quantum system of this form will be physically realizable which means that the commutation relation (\ref{comm_1}) will hold for all times $t \geq 0$.

 We now describe a single quantum Harmonic oscillator system  which will correspond to the quantum observer; see also \cite{JNP1,GJ09,ZJ11}. 
This system is described by a differential equation of the form
\begin{eqnarray}
\dot x_o(t) &=& A_ox_o(t);\nonumber \\
z_o(t) &=& C_ox_o(t)
 \label{observer}
\end{eqnarray}
where the observer output $z_o$ is the observer estimate variable and $C_o\in \rbb^{1\times 2}$. Also, $A_o \in \rbb^{3
\times 3}$, and $ x_o(t) = [\begin{array}{ccc} q(t) & 
p(t)
\end{array}]\trp$ is a vector of self-adjoint 
non-commutative system variables with $q(t)$ being the position operator and $p(t)$ being the momentum operator; e.g., see \cite{JNP1}. We assume that the plant variables commute with the observer variables. The system dynamics (\ref{observer}) are determined by the system Hamiltonian which is a 
which is a self-adjoint operator on the underlying infinite dimensional Hilbert space for the system $\mathfrak{H}_o$. For the single quantum Harmonic oscillator system under consideration, the system Hamiltonian is determined by the
quadratic form
$\mathcal{H}_o=\half x(0)\trp R_o x(0)$, where $R_o$ is a real symmetric matrix. Then, the corresponding matrix $A_o$ in 
(\ref{observer}) is given by 
\begin{equation}
A_o=2J R_o \label{eq_coef_cond_A}
\end{equation}
 where $J$ denotes the real skew-symmetric $2\times 2$ matrix
$$
J= \left[ \begin{array}{cc} 0 & 1 \\ -1 & 0
\end{array} \right];$$
e.g., see \cite{JNP1}.
The  system variables $x_o(t)$ 
will then satisfy the {\em commutation relations}
\begin{equation}
\label{CCR}
[x_o(t),x_o(t)^T]=2{\pmb i}J \ \mbox{for all } t\geq 0.
\end{equation}
That is, the system will be \emph{physically realizable}; e.g., see \cite{JNP1}.

\begin{remark}
\label{R1}
Note that that the Hamiltonian $\mathcal{H}_o$ is preserved in time for the system (\ref{observer}). Indeed,
$ \mathcal{\dot H}_o = x_o^TR_o\dot x_o = 2 x_o^T R_o J R_o x = 0$ since $R_o$ is symmetric and $J$ is skew-symmetric. 
\end{remark}

\section{Direct Coupling Coherent Quantum Observers}
In our proposed direct coupling coherent quantum observer, the quantum plant (\ref{plant}) will be directly coupled to the coherent quantum observer (\ref{observer}) by introducing a coupling Hamiltonian
\begin{equation}
\label{coupling_hamiltonian}
\mathcal{H}_c=\half x_p(0)\trp R_c x_o(0)+\half x_o(0)\trp R_c\trp x_p(0)
\end{equation}
where $R_c\in \rbb^{3\times 2}$.
The augmented quantum linear system consisting of the quantum plant and the direct coupled  quantum observer is then a quantum system  described by the total Hamiltonian
\begin{eqnarray}
\mathcal{H}_a &=& \mathcal{H}_p+\mathcal{H}_c+\mathcal{H}_o\nonumber \\
&=& r_p^T x_p(0) + \half x_p(0)\trp R_c x_o(0)+\half x_o(0)\trp R_c\trp x_p(0)\nonumber \\
&&+ \half x_o(0)\trp R_o x_o(0)
\label{total_hamiltonian}
\end{eqnarray}
 Then, it follows that the augmented quantum  system is described by the equations
\begin{eqnarray}
\dot x_p(t) &=& -{\pmb i}[x_p(t),\mathcal{H}_a];~ x_p(0)=x_{0p};\nonumber \\
\dot x_o(t) &=& -{\pmb i}[x_o(t),\mathcal{H}_a];~ x_o(0)=x_{0o};\nonumber \\
z_p(t) &=& C_px_p(t);\nonumber \\
z_o(t) &=& C_ox_o(t);
\label{augmented_system}
\end{eqnarray}
e.g., see \cite{EMPUJ1a,EMPUJ4a}.

We now formally define the notion of a direct coupled linear quantum observer.

\begin{definition}
\label{D1}
The matrices $R_o \in \rbb^{2
\times 2}$,  $R_c \in \rbb^{3 \times 2}$, $C_o\in \rbb^{1 \times 2}$ define a {\em direct coupled linear quantum observer} for the quantum plant (\ref{plant}) if the corresponding augmented quantum system (\ref{augmented_system}) is such that
\begin{equation}
\label{average_convergence}
\lim_{T \rightarrow \infty} \frac{1}{T}\int_{0}^{T}(z_p(t) - z_o(t))dt = 0.
\end{equation}
\end{definition}

\section{Constructing a Direct Coupling Coherent Quantum Observer}
We now describe the construction of a direct coupled linear quantum observer.  In this section, we assume that  $A_p =0$ in (\ref{plant}). This corresponds to $r_p = 0$ in the plant Hamiltonian. It follows from (\ref{plant}) that the plant system variables $x_p(t)$ will remain fixed if the plant is not coupled to the observer. However, when the plant is coupled to the quantum observer this will no longer be the case. We will show that if the quantum observer is suitably designed, the plant quantity to be estimated  $z_p(t)$ will remain fixed and the condition (\ref{average_convergence}) will be satisfied. 

We also assume that the matrix $R_c$ is of the form  $R_c = \alpha\beta^T$ where $\alpha = C_p^T \in \rbb^{3}$ and $\beta \in \rbb^{2}$. Then, the total Hamiltonian (\ref{total_hamiltonian}) will be given by 
\[
\mathcal{H}_a = \alpha^Tx_p(0)\beta^Tx_o(0)+ \half x_o(0)\trp R_o x_o(0)
\]
since in this case the quantities $\alpha^Tx_p(0)$ and $\beta^Tx_o(0)$ are commuting scalar operators. 

Now using a similar calculation as in (\ref{Ap}), we calculate 
\begin{eqnarray}
\label{xpt}
\dot x_p(t) &=& -{\pmb i}[x_p(t),\mathcal{H}_a]\nonumber \\
&=&-2\Theta(\alpha)x_p(t)\beta^Tx_o(t). 
\end{eqnarray}
Also to calculate $\dot x_o(t)$, we first observe that
\begin{eqnarray*}
\left[\beta^Tx_o(t),x_o(t)\right] &=& \beta^Tx_o(t)x_o(t)-x_o(t)\beta^Tx_o(t) \nonumber  \\
&=& \left(\beta^Tx_o(t)x_o(t)^T\right)^T-x_o(t)x_o(t)^T\beta\nonumber\\
&=&\left(x_o(t)x_o(t)^T\right)^T\beta-x_o(t)x_o(t)^T\beta\nonumber\\
&=&-\left[x_o(t),x_o(t)^T\right]\beta\nonumber\\
&=& -2 {\pmb i} J \beta 
\end{eqnarray*}
using (\ref{CCR}). Hence, using this result and a similar approach to the derivation of (\ref{eq_coef_cond_A}) in \cite{JNP1}, we obtain 
\begin{eqnarray}
\label{xot}
\dot x_o(t) &=& {\pmb i}[\mathcal{H}_a,x_o(t)]\nonumber \\
&=&{\pmb i}\alpha^Tx_p(t)\left(-2 {\pmb i} J \beta \right)+2JR_ox_o(t)\nonumber \\
&=&2J\beta\alpha^Tx_p(t)+2JR_ox_o(t).
\end{eqnarray}

It follows from (\ref{xpt}) and (\ref{xot}) that the quantity $z_p(t) = C_px_p(t)$  satisfies the differential equation
\begin{eqnarray}
\label{zop}
\dot z_p(t) &=& -2C_p\Theta(\alpha)x_p(t)\beta^Tx_o(t)\nonumber \\
 &=& -2\alpha^T\Theta(\alpha)x_p(t)\beta^Tx_o(t) = 0
\end{eqnarray}
using (\ref{eq:Theta_beta_beta}) and the fact that $\Theta(\alpha)$ is skew symmetric. That is, the quantity $z_p(t)$ remains constant and is not affected by the coupling to the coherent quantum observer:
\[
z_p(t) = z_p(0)~ \forall t \geq 0.
\]
Now using this result in (\ref{xot}), it follows that 
\begin{eqnarray}
\label{xot1}
\dot x_o(t) &=&2J\beta z_p(0)+2JR_ox_o(t).
\end{eqnarray}
Hence, we can write
\begin{eqnarray}
\label{xot2}
\lefteqn{x_o(t)}\nonumber \\
 &=&e^{2JR_ot}x_o(0)+2\int_0^te^{2JR_o(t-\tau)}d\tau J\beta z_p(0)\nonumber \\
&=& e^{2JR_ot}x_o(0)-e^{2JR_ot}\left(e^{-2JR_ot}-I\right)R_o^{-1}\beta z_p(0)\nonumber \\
&=& e^{2JR_ot}\left(x_o(0)+R_o^{-1}\beta z_p(0)\right)-R_o^{-1}\beta z_p(0).
\end{eqnarray}

At this point, we observe that the differential equations (\ref{zop}) and (\ref{xot1}) defining the variables $z_p(t)$ and $x_o(t)$ are linear and closed. That is, we can write
\begin{eqnarray}
\left[\begin{array}{c}\dot z_p(t)\\ \dot x_o(t) \end{array}\right]
&=& \left[\begin{array}{cc}0 & 0 \\ 2J\beta & 2JR_o \end{array}\right]
\left[\begin{array}{c} z_p(t)\\  x_o(t) \end{array}\right].
\label{augmented_system1}
\end{eqnarray}
However, the differential equation (\ref{xpt}) defining the complete vector of plant variables $x_p(t)$ is nonlinear. 

We now choose the parameters of the quantum observer so that  $R_o > 0$ and $C_oR_o^{-1}\beta = -1$. It follows from (\ref{xot2})  that the quantity $z_o(t) = C_ox_o(t)$ is given by
\begin{eqnarray}
\label{zot}
z_o(t) &=& C_oe^{2JR_ot}\left(x_o(0)+R_o^{-1}\beta z_p(0)\right)\nonumber \\
&&-C_oR_o^{-1}\beta z_p(0)\nonumber \\
&=& z_p(0)+ C_oe^{2JR_ot}\left(x_o(0)+R_o^{-1}\beta z_p(0)\right).\nonumber \\
\end{eqnarray}

We now verify that the condition (\ref{average_convergence}) is satisfied for this quantum observer. We recall from Remark \ref{R1} that the quantity $\half x(t)\trp R_o x(t)$
remains constant in time for the linear system:
\[
\dot x(t) = 2JR_o x(t);\quad x(0) = x_0.
\]
That is 
\begin{equation}
\label{Roconst}
\half x(t) \trp R_o x(t) = \half x_0 \trp R_o x_0 \quad \forall t \geq 0.
\end{equation}
However, $x(t) = e^{2JR_ot}x_0$ and $R_o > 0$. Therefore, it follows from (\ref{Roconst}) that
\[
\|e^{2JR_ot}x_0\| \leq \sqrt{\frac{\lambda_{max}(R_o)}{\lambda_{min}(R_o)}}\|x_0\|
\]
for all $x_0$ and $t \geq 0$. Hence, 
\begin{equation}
\label{exp_bound}
\|e^{2JR_ot}\| \leq \sqrt{\frac{\lambda_{max}(R_o)}{\lambda_{min}(R_o)}}
\end{equation}
for all $t \geq 0$.

Now since $J$ and $R_o$ are both non-singular,
\[
\int_0^Te^{2JR_ot}dt = \half e^{2JR_oT}R_o^{-1}J^{-1} - \half R_o^{-1}J^{-1}
\]
and therefore, it follows from (\ref{exp_bound}) that
\begin{eqnarray*}
\lefteqn{\frac{1}{T} \|\int_0^Te^{2JR_ot}dt\|}\nonumber \\
 &=& \frac{1}{T} \|\frac{1}{2}e^{2JR_oT}R_o^{-1}J^{-1} - \frac{1}{2}R_o^{-1}J^{-1}\|\nonumber \\
&\leq& \frac{1}{2T}\|e^{2JR_oT}\|\|R_o^{-1}J^{-1}\| \nonumber \\
&&+ \frac{1}{2T}\|R_o^{-1}J^{-1}\|\nonumber \\
&\leq&\frac{1}{2T}\sqrt{\frac{\lambda_{max}(R_o)}{\lambda_{min}(R_o)}}\|R_o^{-1}J^{-1}\|\nonumber \\
&&+\frac{1}{2T}\|R_o^{-1}J^{-1}\|\nonumber \\
&\rightarrow & 0 
\end{eqnarray*}
as $T \rightarrow \infty$. Hence, (\ref{zot}) implies
\[
\lim_{T \rightarrow \infty} \frac{1}{T}\int_{0}^{T} z_o(t)dt = z_p(0).
\]
Also, (\ref{zop}) implies 
\[
\lim_{T \rightarrow \infty} \frac{1}{T}\int_{0}^{T} z_p(t)dt = z_p(0).
\]
Therefore, condition (\ref{average_convergence}) is satisfied. Thus, we have established the following theorem.

\begin{theorem}
\label{T1}
Consider a quantum plant of the form (\ref{plant}) where  $A_p = 0$. Then the matrices $R_o$, $R_c$, $C_o$ will define direct coupled quantum observer (\ref{observer}) for this quantum plant if the  matrix $R_c$ is of the form  $R_c = \alpha\beta^T$ where $\alpha = C_p^T \in \rbb^{3}$, $\beta \in \rbb^{2}$, $R_o > 0$  and $C_oR_o^{-1}\beta = -1$.
\end{theorem}

We now construct the solution to the differential equation (\ref{xpt}) defining the  vector of plant variables $x_p(t)$. In particular, we wish to write down an expression for the remaining variables in $x_p(t)$ apart from $z_p(t)$. For simplicity, we assume $\alpha^T\alpha = 1$ and construct a matrix $D \in \rbb^{3\times 2}$ such that $\alpha^TD = 0$ and $D^TD=I$. It follows that
\[
\left[\begin{array}{c}\alpha^T\\D^T \end{array}\right]\left[\begin{array}{cc}\alpha & D \end{array}\right]
= \left[\begin{array}{cc} I & 0 \\0 & I\end{array}\right]
\]
and hence 
\[
\left[\begin{array}{cc}\alpha & D \end{array}\right] = \left[\begin{array}{c}\alpha^T\\D^T \end{array}\right]^{-1}.
\]
Now define $w_p(t) = D^Tx_p(t)$ which represents the remaining variables in $x_p(t)$ apart from $z_p(t)$. Then, we have
\[
\left[\begin{array}{c}z_p(t)\\w_p(t) \end{array}\right]= \left[\begin{array}{c}\alpha^T\\D^T \end{array}\right]x_p(t)
\]
and hence
\[
x_p(t) = \left[\begin{array}{cc}\alpha & D \end{array}\right]\left[\begin{array}{c}z_p(t)\\w_p(t) \end{array}\right].
\]
We now use (\ref{xpt}) to obtain
\begin{eqnarray}
\label{wpt}
\lefteqn{\dot w_p(t)}\nonumber \\
&=&-2D^T\Theta(\alpha)\left[\begin{array}{cc}\alpha & D \end{array}\right]\left[\begin{array}{c}z_p(t)\\w_p(t) \end{array}\right]\beta^Tx_o(t) 
\nonumber \\
&=& -2\left(D^T\Theta(\alpha)\alpha z_p(t)+ D^T\Theta(\alpha)Dw_p(t)\right)\beta^Tx_o(t) \nonumber \\
&=& -2D^T\Theta(\alpha)Dw_p(t)\beta^Tx_o(t) 
\end{eqnarray}
using (\ref{eq:Theta_beta_beta}). Now define $A_w = -2D^T\Theta(\alpha)D \in \rbb^{2\times 2}$ and the scalar operator $y_o(t) = \beta^Tx_o(t)$. It follows from (\ref{xot2}) that we can write
\begin{equation}
\label{yot}
y_o(t) = -\beta^TR_o^{-1}\beta x_p(0)+ \beta^Te^{2JR_ot}\left(x_o(0)+R_o^{-1}\beta z_p(0)\right)
\end{equation}
and (\ref{wpt}) becomes
\begin{equation}
\label{wpdot}
\dot w_p(t) = y_o(t)A_w w_p(t)
\end{equation}
since $y_o(t)$ is a scalar operator which commutes with $w_p(t)$. Also, since we have a closed form expression (\ref{yot}) for $y_o(t)$, (\ref{wpdot}) can be regarded as a time varying linear differential equation. Then, we can write the solution to this equation in the form 
\begin{equation}
\label{wp1}
w_p(t) = \Phi(t,0)w_p(0)
\end{equation}
where the transition matrix $\Phi(t,0)$ satisfies the differential equation
\[
\frac{d\Phi(t,0)}{dt} = y_o(t)A_w \Phi(t,0);~~\Phi(0,0)=I;
\]
e.g., see Chapter 3 of \cite{RUG96}. Furthermore, we can write down an expression for  $\Phi(t,0)$ using the Peano-Baker series:
\begin{eqnarray*}
\lefteqn{\Phi(t,0)=}\nonumber \\
 && I + \int_0^ty_o(\tau_1)A_wd\tau_1 \nonumber \\
&&+ \int_0^ty_o(\tau_1)A_w\int_0^{\tau_1}y_o(\tau_2)A_wd\tau_2d\tau_1 \nonumber \\
&&+ \int_0^ty_o(\tau_1)A_w\int_0^{\tau_1}y_o(\tau_2)A_w\int_0^{\tau_2}y_o(\tau_3)A_wd\tau_3d\tau_2d\tau_1\nonumber \\
&&+ \ldots \nonumber \\
 &=& I + \int_0^ty_o(\tau_1)d\tau_1 A_w\nonumber \\
&&+ \int_0^ty_o(\tau_1)\int_0^{\tau_1}y_o(\tau_2)d\tau_2d\tau_1A_w^2 \nonumber \\
&&+ \int_0^ty_o(\tau_1)\int_0^{\tau_1}y_o(\tau_2)\int_0^{\tau_2}y_o(\tau_3)d\tau_3d\tau_2d\tau_1A_w^3\nonumber \\
&&+ \ldots;  \nonumber \\
\end{eqnarray*}
e.g., see \cite{RUG96}. 
However, as in Example 3.6 in \cite{RUG96}, we can write
\begin{eqnarray*}
\lefteqn{\int_0^ty_o(\tau_1)\int_0^{\tau_1}y_o(\tau_2)\ldots\int_0^{\tau_j}y_o(\tau_{j+1})d\tau_{j+1}d\tau_j\ldots d\tau_1} \nonumber \\
&=& \frac{1}{(j+1)!}\left[\int_0^ty_o(\tau)d\tau\right]^{j+1}.\hspace{3cm}
\end{eqnarray*}
Hence, 
\begin{eqnarray}
\label{Phit}
\Phi(t,0)&=&I + \int_0^ty_o(\tau)d\tau A_w\nonumber \\
&&+\frac{1}{2!}\left(\int_0^ty_o(\tau)d\tau\right)^2A_w^2\nonumber \\
&&+\frac{1}{3!}\left(\int_0^ty_o(\tau)d\tau\right)^3A_w^3+ \ldots \nonumber \\
&=& e^{\int_0^ty_o(\tau)d\tau A_w}.
\end{eqnarray}
Also using (\ref{yot}), we calculate
\begin{eqnarray*}
\lefteqn{\int_0^ty_o(\tau)d\tau}\nonumber \\
 &=& -\beta^TR_o^{-1}\beta x_p(0)t\nonumber \\
&&+ \frac{\beta^T}{2}\left(e^{2JR_ot}-I\right)R_o^{-1}J^{-1}\left(x_o(0)+R_o^{-1}\beta z_p(0)\right). 
\end{eqnarray*}
Hence using (\ref{wp1}) and (\ref{Phit}), we obtain the following closed form expression for $w_p(t)$
\begin{equation}
\label{wpt1}
w_p(t) = e^{\left(\begin{array}{c}2\beta^TR_o^{-1}\beta x_p(0)t\\
-\beta^T\left(e^{2JR_ot}-I\right)R_o^{-1}J^{-1}\\
\times\left(x_o(0)+R_o^{-1}\beta z_p(0)\right)\end{array}\right)D^T\Theta(\alpha)D}w_p(0).
\end{equation}
This expression is a nonlinear function of the vectors of operators $x_p(0)$ and $x_o(0)$.

\begin{remark}
\label{R2}
 We consider the above result for the  case  in which $C_p = [1~0~0]$. This means that the variable to be estimated by the quantum observer is the  first spin operator $\sigma_1(t)$ of the quantum plant; i.e., $z_p(t) = \sigma_1(t)$. By choosing $R_o = I$, $C_o = [1~0]$, $\beta = \left[\begin{array}{l}-1\\0\end{array}\right]$, $\alpha = \left[\begin{array}{l}1\\0\\0\end{array}\right]$ and $D=\left[\begin{array}{ll}0 & 0\\1 & 0\\0 & 1\end{array}\right]$, the conditions of  Theorem \ref{T1} will be satisfied and the observer output variable will be the position operator of the quantum observer $q(t)$; i.e., $z_o(t) = q(t)$.  Before the quantum observer is connected to the quantum plant, the quantities $\sigma_1(t)$, $\sigma_2(t)$ and $\sigma_3(t)$ will remain constant since we have assumed that $A_p = 0$. Now suppose that the quantum observer is connected to the quantum plant at time $t = 0$. According to (\ref{zop}), the plant variable $\sigma_1(t)$ will remain constant at its initial value $\sigma_1(t)=\sigma_1(0)$ but the other plant variables $\sigma_2(t)$ and $\sigma_3(t)$ will evolve in a time-varying and oscillatory way as defined by (\ref{wpt1}). In addition, the observer position operator $q(t)$ will evolve in an oscillatory way as defined by (\ref{zop}) but its time average will converge to $\sigma_1(0)$ according to (\ref{average_convergence}). 

Now suppose that after a sufficiently long time $T$ such that the time average of $q(t)$ has essentially converged to $\sigma_1(0)$, the observer is disconnected from the quantum plant. Then, the plant  operator $\sigma_1(t)$ will remain constant at $\sigma_1(t)=\sigma_1(0)$ and the plant  operators $\sigma_2(t)$, $\sigma_3(t)$ will remain constant at the values $\sigma_2(T)$, $\sigma_2(T)$ respectively  which are determined by the formula (\ref{wpt1}) in terms of $x_p(0)$, $x_o(0)$ and the time $T$. This will be an essentially ``random'' value. If at a later time an observer with the same parameters as above is connected to the quantum plant, then time average of its output $z_o(t) = q(t)$ will again converge to $\sigma_1(0)$ and $\sigma_1(t)$ will remain constant at  $\sigma_1(0)$. However, suppose that instead an observer with different parameters 
$R_o = I$, $C_o = [0~1]$, $\beta = \left[\begin{array}{l}0\\-1\end{array}\right]$, $\alpha = \left[\begin{array}{l}0\\1\\0\end{array}\right]$ and $D=\left[\begin{array}{ll}1 & 0\\0 & 0\\0 & 1\end{array}\right]$
is used.
This observer is designed so that the time average of the observer output $z_o(t) = p(t)$ converges to the  operator $\sigma_2(t)$ of the quantum plant. This quantity is the essentially random value $\sigma_2(T)$ mentioned above. In addition, the previously constant value of $\sigma_1(t)=\sigma_1(0)$ will now be destroyed and will evolve to another essentially random value. This behavior of the quantum observer is similar to the behavior of quantum measurements; e.g., see \cite{WM10}. This is not surprising since the behavior of the direct coupled quantum observers considered in this paper and the behavior of quantum measurements are both determined by the quantum commutation relations which are fundamental to the theory of quantum mechanics.
\end{remark}

\section{Illustrative Example}
\begin{figure}[htbp]
\begin{center}
\includegraphics[width=7cm]{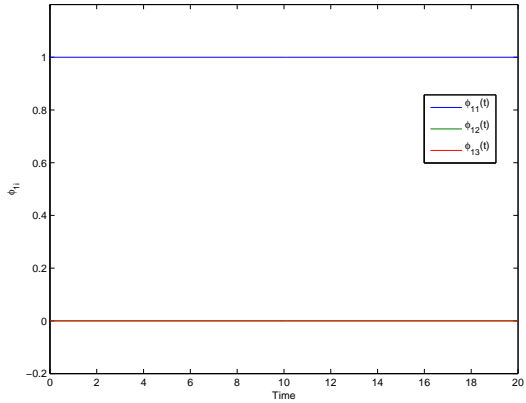}
\end{center}
\caption{Coefficient functions defining $\sigma_1(t)$.}
\label{F3}
\end{figure}

We now present some numerical simulations to illustrate the direct coupled quantum observer described in the previous section. We consider the quantum observer considered in Remark \ref{R2} above where $R_o = I$, $C_o = [1~0]$, $\beta = \left[\begin{array}{l}-1\\0\end{array}\right]$, $\alpha = \left[\begin{array}{l}1\\0\\0\end{array}\right]$ and $D=\left[\begin{array}{ll}0 & 0\\1 & 0\\0 & 1\end{array}\right]$. As described in Remark \ref{R2}, the variable to be estimated by the quantum observer is the  first spin operator $\sigma_1(t)$ of the quantum plant; i.e., $z_p(t) = \sigma_1(t)$. Also, the observer output variable will be the position operator of the quantum observer $q(t)$; i.e., $z_o(t) = q(t)$ where 
$x_o(t) = \left[\begin{array}{l}q(t)\\p(t)\end{array}\right]$. Then the augmented plant-observer system (\ref{augmented_system1}) can be described by the equations
\[
\left[\begin{array}{l}\dot \sigma_1(t)\\ \dot q(t)\\\dot p(t)\end{array}\right]
= A_a \left[\begin{array}{l} \sigma_1(t)\\  q(t)\\ p(t)\end{array}\right]
\]
where
\[
A_a = \left[\begin{array}{ll}0 & 0 \\2 J\beta  & 2 JR_o\end{array}\right]
= \left[\begin{array}{lll}      
     0  &   0  &   0\\
     0  &   0  &   2\\
     2  &  -2  &   0
\end{array}\right].
\]
Then, we can write
\[
\left[\begin{array}{l} \sigma_1(t)\\  q(t)\\ p(t)\end{array}\right] 
= \Phi(t) \left[\begin{array}{l} \sigma_1(0) \\  q(0)\\ p(0)\end{array}\right]
\]
where 
\[
\Phi(t) = \left[\begin{array}{lll}      \phi_{11}(t)  &   \phi_{12}(t)  &  \phi_{13}(t)   \\
     \phi_{21}(t)  &   \phi_{22}(t)  &  \phi_{23}(t)   \\
\phi_{31}(t)  &   \phi_{32}(t)  &  \phi_{33}(t)  
\end{array}\right]
= e^{A_a t}.
\]
Thus, the plant variable to be estimated $\sigma_1(t)$ is given by
\[
\sigma_1(t) = \phi_{11}(t)\sigma_1(0)+\phi_{12}(t)q(0)+\phi_{13}(t)p(t)
\]
and we plot the functions $\phi_{11}(t)$, $\phi_{12}(t)$, $\phi_{13}(t)$ in Figure \ref{F3}. 
From this figure, we can see that $\phi_{11}(t)\equiv 1$, $\phi_{12}(t)\equiv 0$, $\phi_{13}(t)\equiv 0$ and $\sigma_1(t)$ will remain constant at $\sigma_1(0)$ for all $t \geq 0$. 

\begin{figure}[htbp]
\begin{center}
\includegraphics[width=7cm]{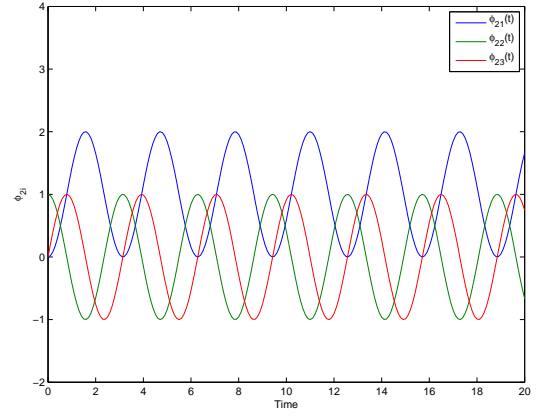}
\end{center}
\caption{Coefficient functions defining $q(t)$.}
\label{F5}
\end{figure}

\begin{figure}[htbp]
\begin{center}
\includegraphics[width=7cm]{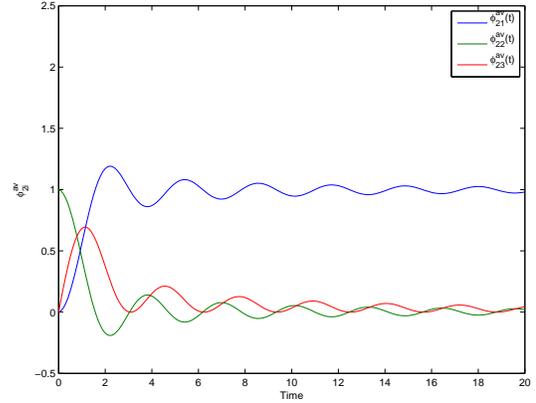}
\end{center}
\caption{Coefficient functions defining the time average of  $q(t)$.}
\label{F6}
\end{figure}

We now consider the output variable of the quantum observer $q(t)$ which is given by
\[
q(t) = \phi_{21}(t)\sigma_1(0)+\phi_{22}(t)q(0)+\phi_{23}(t)p(t)
\]
and we plot the functions $\phi_{21}(t)$, $\phi_{22}(t)$, $\phi_{23}(t)$ in Figure \ref{F5}.
To illustrate the time average convergence property of the quantum observer (\ref{average_convergence}), we now plot the average quantities
\begin{eqnarray*}
\phi_{21}^{ave}(T) &=& \frac{1}{T}\int_0^T\phi_{21}(t)dt\nonumber \\ 
\phi_{22}^{ave}(T) &=& \frac{1}{T}\int_0^T\phi_{22}(t)dt\nonumber \\ 
\phi_{23}^{ave}(T) &=& \frac{1}{T}\int_0^T\phi_{23}(t)dt\nonumber \\ 
\phi_{24}^{ave}(T) &=& \frac{1}{T}\int_0^T\phi_{24}(t)dt\nonumber \\ 
\end{eqnarray*}
in Figure \ref{F6}. 
From this figure, we can see that the time average of $q(t)$  converges to $\sigma_1(0)$ as $t \rightarrow \infty$. 

\begin{figure}[htbp]
\begin{center}
\includegraphics[width=7cm]{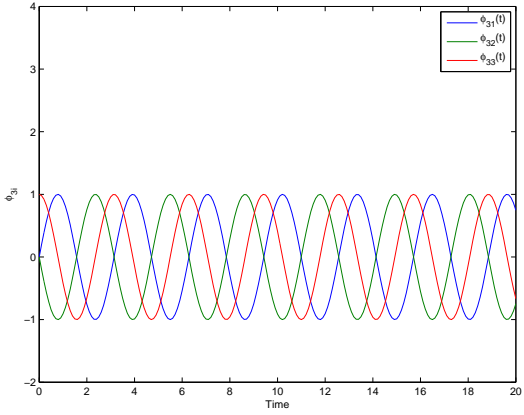}
\end{center}
\caption{Coefficient functions defining $p(t)$.}
\label{F6a}
\end{figure}

We now consider the other variable of the quantum observer $p(t)$ which is given by
\[
p(t) = \phi_{31}(t)\sigma_1(0)+\phi_{32}(t)q(0)+\phi_{33}(t)p(t)
\]
and we plot the functions $\phi_{31}(t)$, $\phi_{32}(t)$, $\phi_{33}(t)$ in Figure \ref{F6a}.

\begin{figure}[htbp]
\begin{center}
\includegraphics[width=7cm]{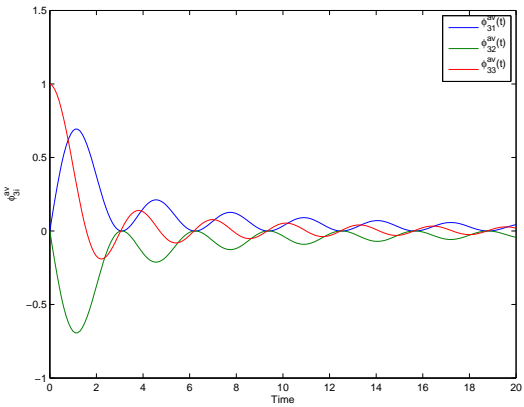}
\end{center}
\caption{Coefficient functions defining the time average of  $p(t)$.}
\label{F6b}
\end{figure}

To investigate the time average  property of the other quantum observer variable, we now plot the average quantities
\begin{eqnarray*}
\phi_{31}^{ave}(T) &=& \frac{1}{T}\int_0^T\phi_{31}(t)dt\nonumber \\ 
\phi_{32}^{ave}(T) &=& \frac{1}{T}\int_0^T\phi_{32}(t)dt\nonumber \\ 
\phi_{33}^{ave}(T) &=& \frac{1}{T}\int_0^T\phi_{33}(t)dt\nonumber \\ 
\end{eqnarray*}
in Figure \ref{F6b}. 

Note that we did not provide numerical simulations for the other plant variables $\sigma_2(t)$ and $\sigma_3(t)$ since the trajectories of these variables are described by the nonlinear relationship (\ref{wpt1}) which is not easily amenable to the type of simulations given above. However, it can be seen from the formula (\ref{wpt1}) that the quantities $\sigma_2(t)$ and $\sigma_3(t)$ will follow complex time-varying oscillatory trajectories. 

\section{Conclusions}
In this paper we have considered a notion of a direct coupling observer for closed quantum systems and given a result which shows how such an observer can be constructed for the case in which the plant is a single spin system and the observer is a single quantum harmonic oscillator. The main result shows the time average convergence properties of the direct coupling observer. We have also presented an illustrative example along with simulations to investigate the behavior of the direct coupling observer. 


\end{document}